\documentclass[aps,showpacs,twocolumn,twoside,amsmath,amssymb,superscriptaddress,prc]{revtex4-2}
\usepackage{graphicx}
\usepackage{tikz}
\usepackage{epstopdf}
\usepackage[percent]{overpic}  
\usepackage{amsmath}
\usepackage{scrextend}  
\usepackage{braket}
\usepackage{bm}
\usepackage{xcolor,xspace}
\usepackage[ulem=normalem]{changes}
\usepackage{hyperref}
\hypersetup{colorlinks=True,urlcolor=blue,linkcolor=blue,citecolor=blue,filecolor=black}

\colorlet{Changes@Color}{red}  
\newcommand{\ba}{\begin{eqnarray}}
\newcommand{\ea}{\end{eqnarray}}
\newcommand{\bsub}{\begin{subequations}}
\newcommand{\esub}{\end{subequations}}
\newcommand\tb{\tilde{\beta}}

\newcommand\+{\dagger}

\begin{document}

\title{
Partial dynamical symmetry from energy density functionals
}

\author{K. Nomura}
\affiliation{Department of Physics, Faculty of Science, University of
Zagreb, HR-10000 Zagreb, Croatia}

\author{N. Gavrielov}\altaffiliation{
  Current address: Center for Theoretical Physics, Sloane Physics Laboratory,
  Yale University, New Haven, CT 06520-8120, USA}
\author{A. Leviatan}
\affiliation{Racah Institute of Physics, The Hebrew University,
Jerusalem 91904, Israel}

\date{\today}

\begin{abstract}
  We show that the notion of partial dynamical symmetry
  is robust and founded on a microscopic many-body theory
  of nuclei.
  Based on the universal energy density functional
  framework, a general quantal boson Hamiltonian is
  derived and shown to have essentially the same
  spectroscopic character as that predicted by the
  partial SU(3) symmetry. The principal conclusion holds
  in two representative classes of energy density
  functionals: nonrelativistic and relativistic.
 The analysis is illustrated in application to the
 axially-deformed nucleus $^{168}$Er. 
\end{abstract}

\keywords{}

\maketitle

\section{Introduction}
Symmetries play a central role in quantum many-body
physics. 
Dynamical symmetry (DS) is a class of symmetry that
appears universally in diverse systems including
hadrons~\cite{Bohm},
nuclei~\cite{IBM}, molecules~\cite{Iachello-Levine}
and atoms~\cite{DS_nano}.
The DS occurs if the Hamiltonian of the system
can be written in terms of Casimir
operators of a chain of nested algebras.
The Hamiltonian is then exactly solvable and 
the spectra and wave functions are completely specified
by the irreducible representations (irreps) of the algebras
in the chain.
In real quantum systems, however, an exact DS rarely occurs.
More often some states obey the patterns required by the
symmetry, but others do not.
This necessitates a certain degree of symmetry-breaking,
a prominent case of which is partial dynamical symmetry
(PDS)~\cite{leviatan2011,leviatan1996}.
Its basic idea
is to relax the stringent conditions imposed by an exact DS
so that solvability and/or good symmetry are
retained by only a subset of states.
Detailed studies employing bosonic and fermionic models
based on spectrum generating algebras, 
have shown that PDSs account quite well
for a wealth of spectroscopic data in various types of
nuclei~\cite{leviatan2011,leviatan1996,leviatan1999,
  casten2014,casten2016,couture2015,leviatan2013,
  ramos2009,vanisacker2015,escher2000,rowe2001,vanisacker2014}
and are relevant to related quantum phase transitions
and shape-coexistence~\cite{leviatan2007,macek2014,
leviatan2016,leviatan2017,leviatan2018}.

One drawback in the implementation of symmetry-based
notions in composite systems is that they are,
in most cases, used
without much considerations of their microscopic basis, 
i.e., connection to more fundamental degrees of freedom. 
With that in mind, the role of an emergent Sp(3,R) DS
in light nuclei has been recently demonstrated
within a symplectic no-core configuration
interaction framework built on realistic nucleon-nucleon
potentials~\cite{dytrych2020,mccoy2020}.
The existence of SU(3) quasi dynamical symmetry 
evolved into a DS by means of the similarity 
renormalization group, has been demonstrated for 
$^{36}$Ar using a universal $sd$ shell model interaction 
\cite{johnson2020}. 
In the same context, and motivated by the
empirical manifestations of partial symmetries in
heavy nuclei, a microscopic justification of PDS
is called for.
In this work, we present a first proof of principle that the notion
of PDS is robust and founded on 
a microscopic quantum many-body theory of nuclei.
This proposition is illustrated in the example of the
Hamiltonian with partial SU(3) symmetry in the framework
of the interacting boson model (IBM)~\cite{IBM},
widely used for describing
collective states in
nuclei in terms of monopole ($s$)
and quadrupole ($d$) bosons.
Based on a fermionic mean-field framework,
we determine microscopically a general
boson Hamiltonian, which is then shown to
produce eigenstates that are
similar in structure to those of the PDS one.
We~apply the procedure to $^{168}$Er,
a typical example of an axially-deformed nucleus,
in which the SU(3)-PDS was
previously recognized on phenomenological
grounds~\cite{leviatan1996,leviatan1999,casten2014}.
\begin{figure*}[t]
\begin{center}
  \begin{tabular}{cc}
\includegraphics[width=0.45\linewidth]{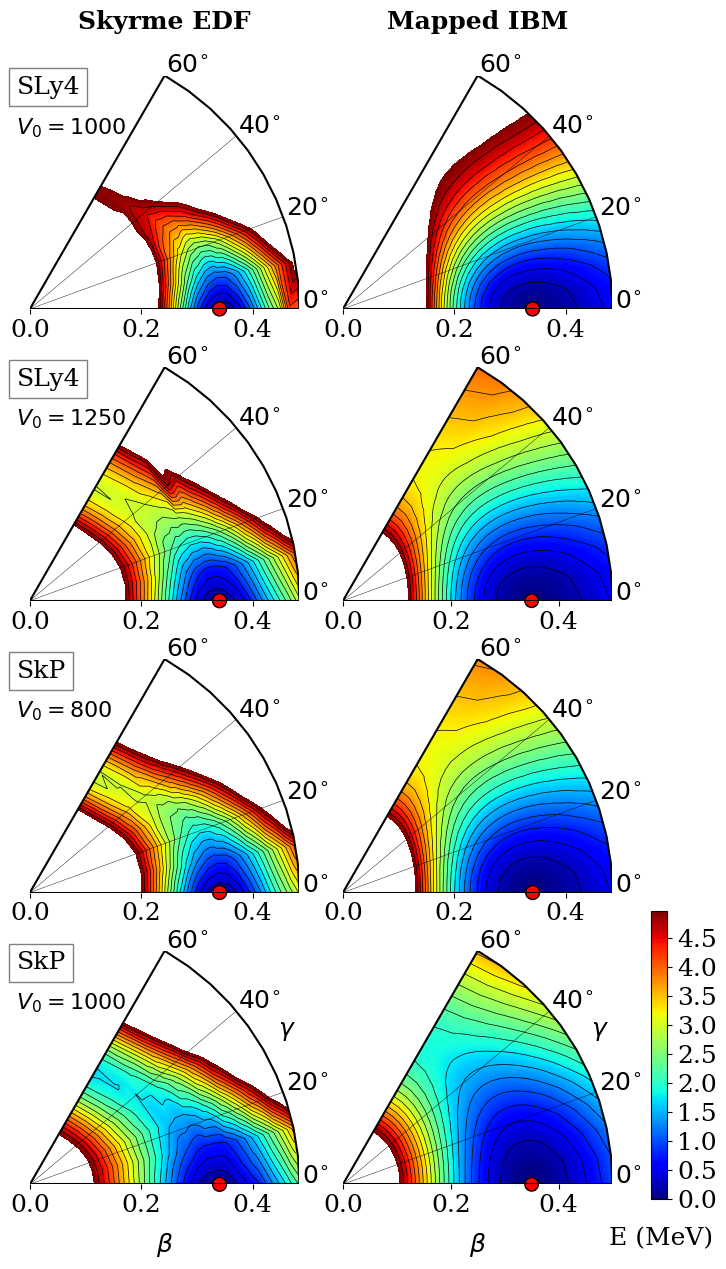} &
\includegraphics[width=0.45\linewidth]{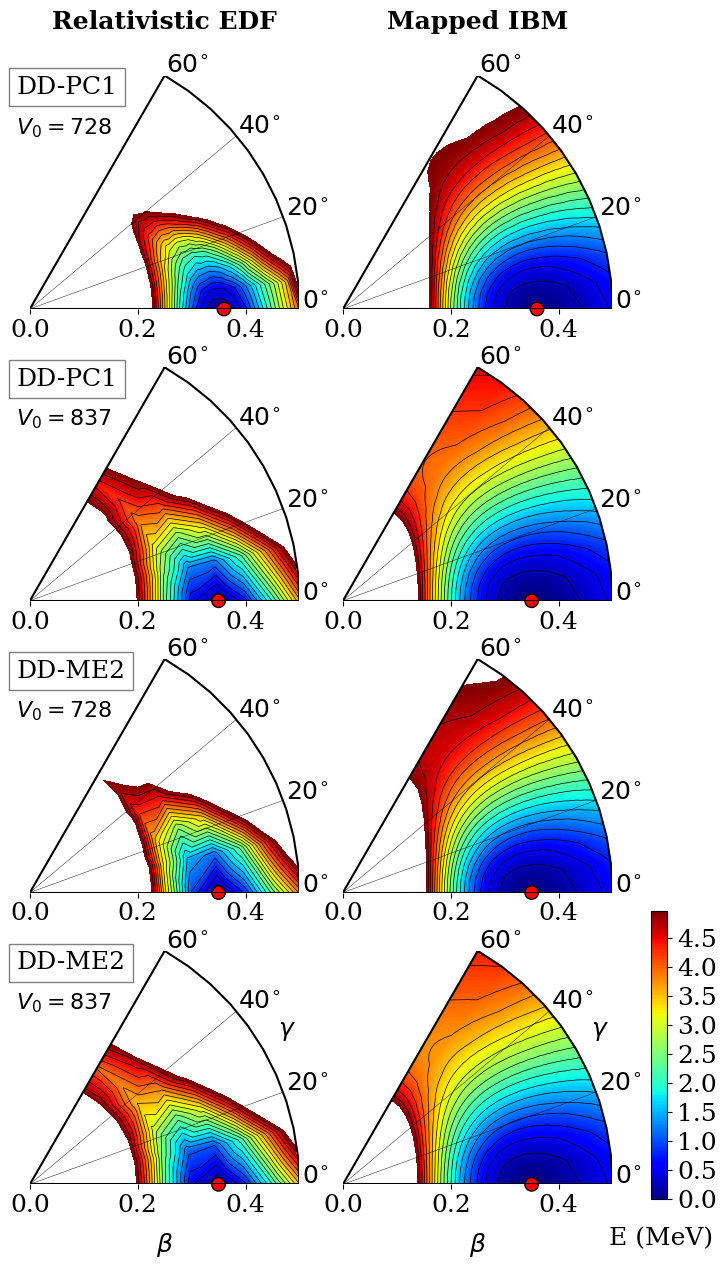} \\
\end{tabular}
\caption{SCMF energy surfaces in the $\beta$-$\gamma$
  plane for $^{168}$Er, based on
  the nonrelativistic Skyrme SLy4 and SkP EDFs
  (first column) and
 the relativistic DD-PC1 and DD-ME2 EDFs (third column)
 with different
 values of pairing strengths $V_{0}$ in units of MeVfm$^{3}$.
 The corresponding mapped IBM energy surfaces 
 are plotted on the second and fourth columns.
Contour spacing is 0.25 MeV, and 
 the global minimum is indicated by a solid circle.
}
\label{fig:pes} 
\end{center}
\end{figure*}

\section{IBM framework and SU(3) PDS}
Shapes and symmetries in nuclei can be
studied in the IBM framework with
the following Hamiltonian~\cite{leviatan1987}
\begin{align}
\label{eq:ham}
\hat{H} = h_{0} P^\+_0(\beta_0)P_{0}(\beta_0) + h_2
P^{\+}_2(\beta_{0})\cdot\tilde{P}_{2}(\beta_0)
+ \rho\hat{L}\cdot\hat{L}. 
\end{align}
Here $P^\+_0(\beta_0) = d^\+\cdot d^\+
- \beta_{0}^{2}(s^{\+})^2$,
$P^\+_{2\mu}(\beta_0) = \beta_0\sqrt{2}d^\+_{\mu}s^\+
+ \sqrt{7}(d^\+d^\+)^{(2)}_{\mu}$,
$\tilde{P}_{2\mu}(\beta_0)=(-1)^{\mu}{P_{2,-\mu}}(\beta_{0})$,
$\hat{L}$ the angular momentum operator and
standard notation of angular momentum coupling is used.
The first two terms in Eq.~(\ref{eq:ham})
comprise the most general intrinsic
Hamiltonian appropriate for the dynamics of
a prolate-deformed shape. This is consistent with
the energy surface obtained by
its expectation value in a coherent (intrinsic)
state~\cite{ginocchio1980,dieperink1980},
which has the form
\begin{align}
\label{eq:pes}
 E_{\mathrm{IBM}}(\tb,\gamma)
&=N(N-1)(1+\tb^2)^{-2}[\,h_0(\tb^2-\beta^2_0)^2
\nonumber \\
&+2h_2\tb^2(\tb^2-2\beta_0\tb\cos{3\gamma}+\beta_0^2)\,].
\end{align}
Here $(\tb,\gamma)$ are the quadrupole shape parameters
in the IBM.
For $h_0,h_2\!\geqslant\! 0$, the surface has a
global minimum at
$(\tb\!=\!\beta_{0}\!>\!0,\gamma\!=\!0^{\circ})$,
corresponding to a prolate-deformed equilibrium shape.
The intrinsic Hamiltonian determines the band-structure.
The last term in Eq.~(\ref{eq:ham}) determines the in-band
rotational splitting and its contribution to the energy surface
is $1/N$ suppressed, hence negligible.
For $\beta_0\!=\!\sqrt{2}$ and $h_0\!=\!h_2$,
the Hamiltonian~(\ref{eq:ham}) involves
the Casimir operators of the algebras in the chain
${\rm U(6)}\supset {\rm SU(3)}\supset {\rm SO(3)}$,
hence exhibits an SU(3) DS. The spectrum consists of SU(3)
multiplets with the states $\ket{[N](\lambda,\mu)KL}$
specified by the total boson number $N$, 
the SU(3) irrep $(\lambda,\mu)$, the angular
momentum $L$, and the label $K$ which
corresponds to the projection of the 
angular momentum on the symmetry axis. 
The lowest multiplets have
$(\lambda,\mu)\!=\!(2N,0)$ which contains
the ground band~$g(K\!=\!0)$,
and $(\lambda,\mu)\!=\!(2N-4,2)$
which contains both
the $\beta(K\!=\!0)$ and $\gamma(K\!=\!2)$ bands.
For $\beta_{0}\!=\!\sqrt{2}$ and $h_{0}\neq h_{2}$, the
SU(3) symmetry is broken but selected bands maintain
it~\cite{leviatan1996}.
In particular, the
$(\lambda,\mu)$ SU(3)-classification remains intact
for the ground and $\gamma$ bands,
but the $\beta$ band is mixed.
By definition, the resulting Hamiltonian has SU(3)-PDS.
Previous studies in $^{168}$Er employed
the Hamiltonian~(\ref{eq:ham}) with $\beta_0\!=\!\sqrt{2}$
and remaining parameters determined from
a fit~\cite{leviatan1996}.
The SU(3)-PDS predictions compared favorably
with the empirical data for energies and E2
rates~\cite{leviatan1996,leviatan1999,casten2014}.
In what follows, we verify whether the features of
SU(3)-PDS are realized by a microscopically-derived
boson Hamiltonian. 

\section{SCMF framework and mapping procedure}
Among contemporary microscopic approaches, the nuclear
energy density
functional (EDF) framework allows for a reliable
quantitative prediction
of ground-state properties and collective excitations of
nuclei over the
entire region of the nuclear chart~\cite{schunk2019}. 
The basic implementation of the EDF framework is in
self-consistent
mean-field (SCMF) methods, in which an EDF is
constructed as a
functional of one-body nucleon density matrices that
correspond to a
single product state. 
Both nonrelativistic~\cite{bender2003,robledo2019} and
relativistic~\cite{vretenar2005,niksic2011} EDFs have
been successfully applied to numerous studies of nuclear
shape-related
phenomena and the resulting excitation modes and decay
properties. 
In the present work, we consider
these two representative classes
of the EDF framework, so as to ensure the robustness
of the results.

The starting point is a set of constrained SCMF calculations of an
energy surface~\cite{RS}. The constraints
refer
to those for mass
quadrupole moments, which are associated with the polar
deformation parameters $\beta$ and $\gamma$~\cite{BM_II}.
$\beta$ is proportional to the intrinsic
quadrupole moment and
$\gamma$
specifies the departure from axiality.
For $\gamma\!=\!0^{\circ}$ ($\gamma\!=\!60^{\circ}$)
the nucleus is prolate (oblate) deformed, and
intermediate values $0^{\circ}<\gamma<60^{\circ}$
correspond to triaxial shapes. 
The calculated
SCMF energy surfaces $E_{\mathrm{SCMF}}(\beta,\gamma)$
for $^{168}$Er, are displayed
on the first and third columns of Fig.~\ref{fig:pes}.
The former surfaces
are obtained using the Hartree-Fock plus BCS
model~\cite{ev8,ev8r} with the SLy4~\cite{SLy4} and
SkP~\cite{SkP} parameterizations of the Skyrme
EDF~\cite{skyrme}, while the latter
using the relativistic Hartree-Bogoliubov
model~\cite{vretenar2005,DIRHB} with the density-dependent
point-coupling (DD-PC1)~\cite{DDPC1} and meson-exchange
(DD-ME2)~\cite{DDME2} functionals. 
Pairing correlations are taken into account by employing
the density-dependent delta force and the separable pairing 
force of finite range \cite{tian2009} in
the Skyrme and relativistic frameworks, respectively.
%
For each EDF we consider
several choices of pairing strength $V_{0}$.
Specifically, for the nonrelativistic Skyrme EDFs:
SLy4 with $V_{0}\!=\!1000$ and $1250$ MeVfm$^{3}$,
SkP with $V_{0}\!=\!800$ and $1000$ MeVfm$^{3}$.
A smooth cut-off~\cite{SkP} 
of 5 MeV below and above the Fermi surface
is invoked for these zero-range
forces.
The cut-off dependence of physical quantities 
was investigated in ~\cite{karatzikos2010}.
For the relativistic EDFs: 
both DD-PC1 and DD-ME2 with
$V_{0}\!=\!728$ and $837$ MeVfm$^{3}$
(an increase by 15 \%).
The separable pairing 
force employed is practically identical to the finite-range 
Gogny interaction D1S, which does not need
a cut-off and is known to give
an excellent 
description of pairing over the periodic table.
For nuclear matter, the pairing gap produced by this force
is in good agreement with the results of ab-initio 
calculations with the Bonn potential~\cite{serra2001}.

As is evident from Fig.~\ref{fig:pes}, all adopted  EDFs
lead to energy surfaces accommodating a pronounced
prolate-deformed
global minimum
$(\beta\!\approx\! 0.35,\gamma\!=\!0^{\circ})$.
The minimum tends to be less steep
in $\beta$ and $\gamma$,
for larger pairing strengths.
This is anticipated because the quadrupole response 
function
has two-quasiparticle energies in the denominator 
which increase with the pairing strength \cite{RS}.
\begin{table}[t]
  \caption{\label{tab:para}
 Parameters $h_{0},\,h_{2},\,\rho$ (in keV) and
$\beta_{0}$, of the Hamiltonian~(\ref{eq:ham})
 obtained from SCMF calculations based on
 nonrelativistic Skyrme SLy4 and SkP EDFs, and
 relativistic DD-PC1 and DD-ME2 EDFs, with
 pairing strengths $V_{0}$ (in MeV fm$^{3}$).
 The corresponding parameters for
 SU(3)-PDS~\cite{leviatan1996},
 are also shown. $E(2_2)$ and $E(0_2)$
 are the calculated
 bandhead energies (in keV) for the $\gamma$ and $\beta$ bands
 and $R=\tfrac{E(0_2)}{E(2_2)}$. For $^{168}$Er,
$E(2_2)\!=\!821,\,E(0_2)\!=\!1217$ (in keV)
and $R\!=\!1.48$~\cite{nndc}.}
\begin{center}
 \begin{ruledtabular}
  \begin{tabular}{lccccc|ccc}
    EDF & $V_{0}$  & $h_0$ & $h_2$ & $\rho$ & $\beta_0$ &
    $E(2_2)$ & $E(0_2)$ & $R$\\
\hline
SLy4 & $1000$ & 10 & 5.3 & 11.8 & 1.59$\;$ & 1132 & 1911 & 1.68 \\
    & $1250$ & 10.4 & 4.0 & 12.3 & 1.39 & 809 & 1334 & 1.65 \\
SkP & $800$  & 10.5 & 3.7 & 12.6 & 1.45 & 776 & 1306 & 1.68 \\
    & $1000$ & 30.6 & 4.4 & 12.2 & 0.99 & 672 & 1087 & 1.62 \\
DD-PC1 & $728$ & 10.5 & 5.1 & 11.74 & 1.59 & 1092 & 1889 & 1.73 \\
       & $837$ & 9.8  & 4.4 & 11.73 & 1.51 & 925 & 1564 & 1.69 \\
DD-ME2 & $728$ & 10.4 & 4.8 & 11.74 & 1.59 & 1032 & 1794 & 1.74 \\
       & $837$ & 9.9 & 4.2 & 11.73 & 1.50 & 883 & 1499 & 1.70\\
\hline
SU(3)-PDS & & 8.0 & 4.0 & 13.0 & $\sqrt{2}$ & 822 & 1220 & 1.48\\
  \end{tabular}
 \end{ruledtabular}
 \end{center}
\end{table}

The IBM Hamiltonian of Eq.~(\ref{eq:ham})
is derived by the methods 
of~\cite{nomura2008,nomura2010,nomura2011rot}.
The parameters $\{h_{0}$, $h_{2}$, $\beta_{0}\}$
are determined by mapping the microscopic energy
surface $E_{\mathrm{SCMF}}(\beta,\gamma)$, obtained
for a given EDF, onto the corresponding IBM surface
$E_{\mathrm{IBM}}(\beta,\gamma)$ of Eq.~(\ref{eq:pes}).
The condition $E_{\mathrm{SCMF}}(\beta,\gamma)\approx
E_{\mathrm{IBM}}(\beta,\gamma)$  is imposed to ensure
similar topology
in the neighborhood of the global minimum.
(The two surfaces are expressed in terms of $\beta$,
since the IBM and SCMF deformations are related by
$\tb=C\beta$, where the constant $C$ is determined by the mapping). 
The parameter $\rho$, Eq.~(\ref{eq:ham}),
is obtained by equating the
cranking moment of inertia in the IBM
to the Thouless-Valatin value \cite{delaroche2010},
the procedure discussed in detail
in~\cite{nomura2011rot}.
The mapped IBM energy surfaces, based on the
nonrelativistic and relativistic EDFs, are
shown on the second and fourth columns of
Fig.~\ref{fig:pes}, respectively.
One clearly sees that the IBM and microscopic surfaces
share common essential features near and
up to a few MeV above the global minimum.
\begin{figure*}[t]
\begin{center}
\includegraphics[width=0.65\linewidth]{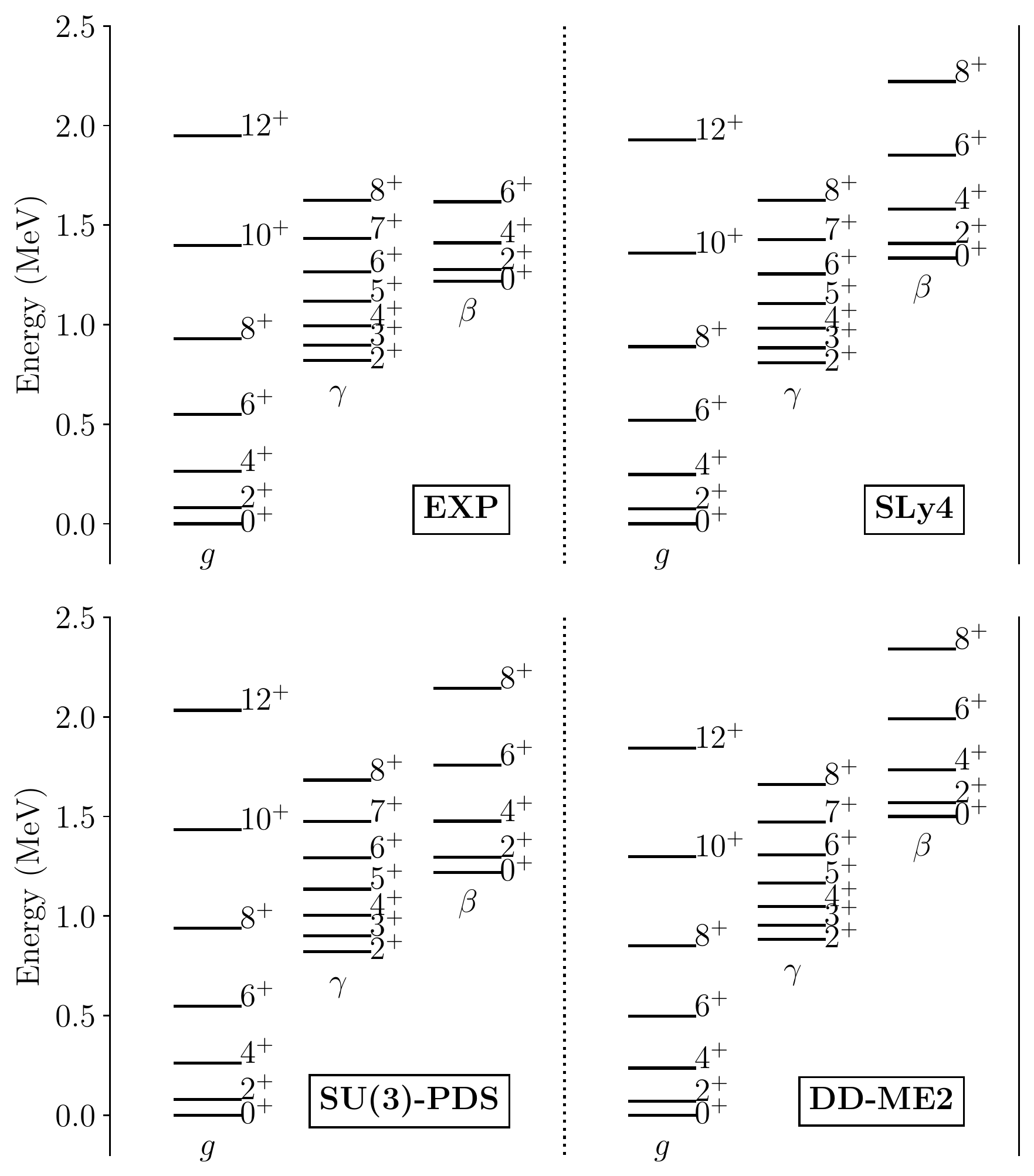}
\caption{\small
Experimental~\cite{nndc} (EXP) and SU(3)-PDS~\cite{leviatan1996}
spectra for $^{168}$Er, compared with the spectra
resulting from EDF-based IBM
calculations for the Skyrme SLy4 EDF with
pairing strength $V_{0}\!=\!1250$ MeVfm$^{3}$, and for
the relativistic EDF DD-ME2 with
$V_{0}\!=\!837$ MeVfm$^{3}$.}
\label{fig:spectra} 
\end{center}
\end{figure*}
\begin{figure*}[t]
\begin{center}
\includegraphics[width=0.75\linewidth]{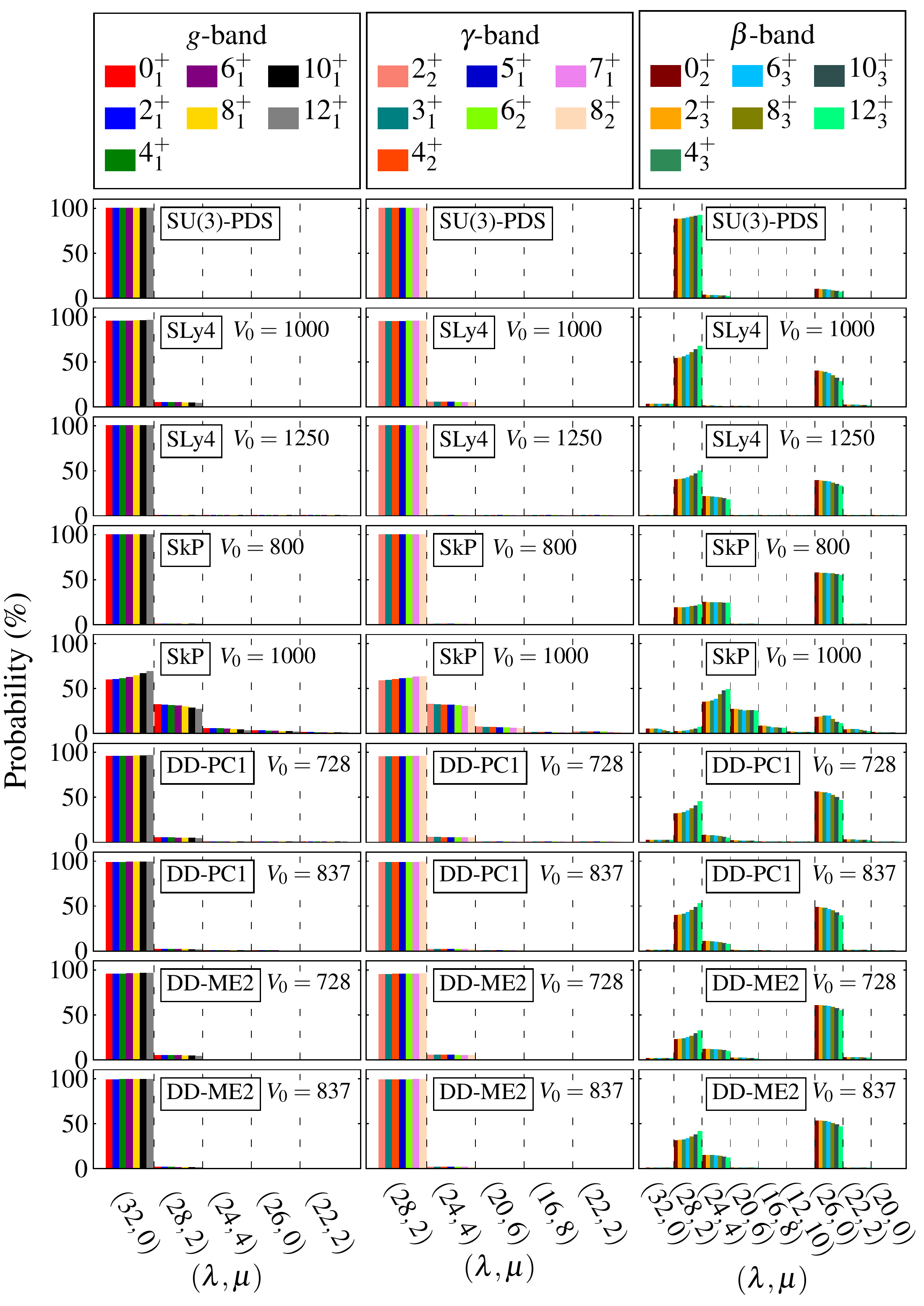} 
\caption{SU(3) $(\lambda,\mu)$-decomposition of states in
  the ground ($g$), $\gamma$ and $\beta$ bands,
  for the SU(3)-PDS and various EDF-based calculations.
  Shown are probabilities larger than 0.5~\%.
  The histograms shown from left-to-right for each band,
  correspond to the $L_i$ states listed in the upper
  panels in the order top-to-bottom left-to-right.}
\label{fig:decom} 
\end{center}
\end{figure*}

\section{EDF-based IBM Hamiltonians and spectra}
The values of the Hamiltonian parameters, derived
microscopically from various EDFs,
are given in Table~\ref{tab:para}, along with
the corresponding SU(3)-PDS parameters (obtained
from a fit~\cite{leviatan1996}).
For SU(3)-PDS, $h_{0}/h_{2}=2$, while in most
SCMF calculations, $1.9<h_{0}/h_{2}<2.8$,
consistent with values obtained in global
IBM fits in
the rare-earth region~\cite{leviatan1999}.
The derived values of $\beta_{0}$ are close
or slightly larger than the SU(3)-PDS value
($\beta_0=\sqrt{2}\approx 1.412$). 
A notable exception are the parameters derived
from the SkP EDF with pairing strength
$V_{0}\!=\!1000$ MeVfm$^{3}$,
which exhibit pronounced large ratio
$h_{0}/h_{2}=6.95$ and small  $\beta_{0}=0.99$.
This is a consequence of the fact that
the corresponding SCMF
energy surface for this case, shown in Fig.~\ref{fig:pes},
is peculiarly soft in the  $\gamma$ deformation, with
a shallow local minimum on
the oblate side. For any chosen EDF,
a larger pairing strength results in a larger
(smaller) value for  $h_{0}/h_{2}$ ($\beta_0$).

Excitation spectra appropriate for $^{168}$Er are
obtained for each EDF by diagonalizing~\cite{arbmodel}
the Hamiltonian~(\ref{eq:ham}) using the parameters
in Table~\ref{tab:para} and $N\!=\!16$.
Typical spectra resulting from representative
nonrelativistic and relativistic EDFs
are displayed in Fig.~\ref{fig:spectra}. They
satisfactorily conform with the calculated SU(3)-PDS
spectrum which, in turn, agrees with experimental
spectrum. The bandhead energies, $E(2_2)$ and $E(0_2)$
for the $\gamma$ and $\beta$ bands, and their ratios for
the different cases, are listed in Table~\ref{tab:para}.
In general, the descriptions for the ground and $\gamma$ 
bands are stable with respect to different choices of EDFs.
The description of the $\beta$-band is more case-sensitive
and all EDFs place $E(0_2)$ above the empirical and
SU(3)-PDS values. The following observations are in order.
(i)~The relativistic EDFs generally result in higher
$\beta$-band energies than the Skyrme EDFs.
(ii)~The increase of the pairing strength ($V_0$)
systematically decreases the $\beta$-band energies.
(iii)~The SkP EDF with $V_{0}\!=\!1000$ MeVfm$^{3}$, is
the only case where both $E(2_2)$ and $E(0_2)$ are
placed below the SU(3)-PDS and empirical values.

\section{Symmetry analysis}
Analysis of wave functions is a more sensitive measure
to quantify the
similarities and differences in structure between
the EDF-based IBM Hamiltonians and SU(3)-PDS.
Fig.~\ref{fig:decom} shows the SU(3)
$(\lambda,\mu)$-decomposition for member states of the
lowest bands in $^{168}$Er.
For SU(3)-PDS, the ground and $\gamma$ bands are pure
with SU(3) character $(2N,0)$ and $(2N-4,2)$, respectively,
whereas the $\beta$ band contains a mixture
of irreps: $(2N-4,2)$ 87.5 \%, $(2N-6,0)$ 9.6 \%, and
$(2N-8,4)$ 2.9 \%, with $N=16$. 
Remarkably, for all nonrelativistic and relativistic EDFs
considered (except SkP with pairing strength
$V_{0}\!=\!1000$ MeVfm$^{3}$),
the mapped IBM Hamiltonians reproduce very well
the SU(3)-PDS prediction of SU(3)-purity for the ground
and $\gamma$ bands, with probability larger than 95\%.
This clearly demonstrates the robustness of the PDS
notion and its microscopic roots.
The structure of the $\beta$ band is more sensitive to 
the choice of EDF.
Its SU(3) mixing is governed by the values of the
parameters $\beta_{0}$ and ratio $h_{0}/h_{2}$ which,
in turn, reflect the different topology of the
corresponding SCMF surfaces.
Although the dominance of the
$(2N-4,2)$, $(2N-6,0)$, and
$(2N-8,4)$ irreps in the $\beta$ band is generally
observed in all cases, their relative weights differ
from those of SU(3)-PDS.
This may indicate that additional degrees of freedom
not included in the IBM ({\it e.g.}, quasi particles)
contribute to the structure of the $K=0_2$ band in
$^{168}$Er.
Again, the situation is different for the EDF
SkP with  $V_{0}\!=\!1000$ MeVfm$^{3}$ for which
the SU(3) decomposition exhibits large fragmentation.
From all the EDFs considered, the SLy4 and SkP with
$V_{0}\!=\!1250$ and $800$ MeVfm$^{3}$, respectively,
appear to yield spectral properties which are
closest to the SU(3)-PDS predictions for $^{168}$Er
(SU(3) purity for the ground and $\gamma$ bands with
probability 99.8\%).

\section{Conclusions}
We have shown that
the occurrence of partial dynamical symmetry (PDS) in
nuclei can be justified from a microscopic point of view.
By employing the constrained mean-field methods with
choices of the universal energy density functionals and
pairing interactions, in combination with symmetry
analysis of the wave functions of the mapped IBM
Hamiltonians, we arrived at an efficient procedure to
test and explain the emergence of PDS in nuclei.
An application to $^{168}$Er, has shown that the
boson Hamiltonians derived from known EDFs
in this region, produced eigenstates whose properties
resemble those of SU(3)-PDS.
The fact that these results are valid for
both nonrelativistic and relativistic EDFs
with several choices of pairing strengths,
highlights the robustness of the PDS notion
and its association with properties of
the multi-nucleon dynamics in nuclei.

The results of the present investigation
pave the way for a number of research avenues.
(i)~Exploring the microscopic origin
of other types of PDSs, {\it e.g.}, SO(6)-PDS in
$\gamma$-soft nuclei~\cite{ramos2009}.
(ii)~When a PDS is found to be manifested empirically in
certain nuclei,
it can be used to constrain,
improve and optimize
({\it e.g.}, choice of the pairing strength)
a given EDF in that region.
(iii)~Exploiting the demonstrated linkage between
the microscopic EDF framework and the algebraic
PDS notion, to predict uncharted regions of exotic
nuclei, awaiting to be explored, where partial symmetries
can play a role.\\

\acknowledgements
The work of K.N.
is supported by the Tenure Track Pilot Programme of 
the Croatian Science Foundation and the 
\'Ecole Polytechnique F\'ed\'erale de Lausanne,
and the Project TTP-2018-07-3554 Exotic Nuclear Structure
and Dynamics, with funds of the Croatian-Swiss Research
Programme and also
by the QuantiXLie Centre of
Excellence, a project co-financed by the Croatian
Government and
European Union through the European Regional Development
Fund -- the Competitiveness and Cohesion Operational
Programme (Grant KK.01.1.1.01.0004).
The work of N.G. and A.L. is supported by the Israel
Science Foundation Grant 586/16.
N.G. acknowledges support by the Israel Academy of Sciences for
a Postdoctoral Fellowship Program in Nuclear Physics.


\end{document}